%% file: femto.tex
\documentclass[ALICE,manyauthors]{cernphprep}
\usepackage{graphics}
\usepackage{multicol}
\usepackage{lineno}
\usepackage{color}
\usepackage{subfigure} 
\usepackage{hyperref}

\newcommand {\ee} {\mbox{$e^+e^-$}~}

\def\qout      {q_{\rm out}}
\def\qside     {q_{\rm side}}
\def\qlong     {q_{\rm long}}

\def\Qout      {$\qout$}
\def\Qside     {$\qside$}
\def\Qlong     {$\qlong$}

\def\rout      {R_{\rm out}}
\def\rside     {R_{\rm side}}
\def\rlong     {R_{\rm long}}
\def\rol       {R_{\rm ol}}

\def\Rout      {$\rout$}
\def\Rside     {$\rside$}
\def\Rlong     {$\rlong$}
\def\Rol       {$\rol$}
\def\tauf      {$\tau_f$}

\def\pbpb      {Pb--Pb}
\def\auau      {Au--Au}
\def\dedx      {${\rm d}E/{\rm d}x$}
\def\pt        {$p_T$}

\def\kt        {$k_T$}
\def\meankt    {$\langle$\kt$\rangle$}
\def\pim       {$\pi^-$}

\def\sqrts     {$\sqrt{s_{\rm NN}}$}
\def\ene       {$\sqrt{s_{\rm NN}}=2.76$~TeV}
\def\dndeta    {${\rm d}N_{\rm ch}/{\rm d}\eta$}

\def\meandndeta{$\langle$\dndeta$\rangle$}
\def\deg       {$^{\rm o}$}
\def\gevc      {~GeV/$c$}
\def\dens      {charged-particle pseudorapidity density}

\begin{document}

\begin{titlepage}
\PHnumber{2010-078}             
\PHdate{17 January 2011}              

\title{Two-pion Bose--Einstein correlations in central \pbpb\ collisions \\ at \ene}
\ShortTitle{}
\Collaboration{ALICE Collaboration%
\thanks{See Appendix~\ref{app:collab} for the list of collaboration members}}
\ShortAuthor{ALICE Collaboration}

\begin{abstract}
  The first measurement of two-pion Bose--Einstein correlations 
  in central \pbpb\ collisions at \ene\ at the Large Hadron Collider 
  is presented. 
  We observe a growing trend with energy now not only for the longitudinal 
  and the outward but also for the sideward pion source radius.  
  The pion homogeneity volume and the decoupling time are significantly 
  larger than those measured at RHIC. 
\end{abstract}
\end{titlepage}
\setcounter{page}{2}


\section{\label{sec:intro}Introduction}

Matter at extremely high energy density created in central collisions 
of heavy ions at the Large Hadron Collider (LHC) is the main object of 
study of ALICE (A Large Ion Collider Experiment)~\cite{Aamodt:2008zz,
Carminati:2004fp,Alessandro:2006yt}. 
Under these conditions the Quark-Gluon Plasma (QGP), a state characterized 
by partonic degrees of freedom, is thought to be formed~\cite{Cabibbo:1975ig, 
Shuryak:1980tp,Heinz:2000bk,Adams:2005dq,Adcox:2004mh,Back:2004je,Arsene:2004fa}. 
The highly compressed strongly-interacting system created in these collisions  
is expected to undergo longitudinal and transverse expansion. 
The first measurement of the elliptic flow in the \pbpb\ system at the LHC 
confirmed the presence of strong collective motion and the hydrodynamic 
behavior of the system~\cite{Aamodt:2010pa}. 
While the hydrodynamic approach is rather successful in describing the 
momentum distributions of hadrons in ultrarelativistic nuclear collisions 
(for recent reviews of hydrodynamic models see Refs.~\cite{Ollitrault:2007du,
Kestin:2008bh,Teaney:2001av,Satarov:2006jq,Huovinen:2006jp}), the spatial 
distributions of decoupling hadrons are more difficult to 
reproduce~\cite{Pratt:2009hu} and thus provide important model constraints 
on the initial temperature and equation of state of the system~\cite{Lisa:2005dd}. 
Experimentally, the expansion rate and the spatial extent at decoupling are 
accessible via intensity interferometry, a technique which exploits the 
Bose--Einstein enhancement of identical bosons emitted close by in phasespace. 
This approach, known as Hanbury Brown--Twiss analysis 
(HBT)~\cite{HBT1,Hanbury:1954wr}, has been successfully applied in 
\ee~\cite{Kittel:2001zw}, hadron--hadron and lepton--hadron~\cite{Alexander:2003ug}, 
and heavy-ion~\cite{Lisa:2005dd} collisions. 

In this Letter, we report on the first measurement of HBT radii for heavy-ion 
collisions at \ene\ at the LHC 
and discuss the space-time properties of the system generated at these record 
energies in the context of systems created at lower energies, measured 
over the past quarter of a century~\cite{Lisa:2005dd}.
Like with such studies at RHIC and SPS energies, our measurements should 
provide strong constraints for models that aspire to describe the dynamic 
evolution of heavy ion collisions at the LHC.

\section{\label{sec:exp}Experiment and data analysis}
The data were collected in 2010 during the first lead beam running period 
of the LHC. 
The runs used in this analysis were taken with beams of either 4 or 66 bunches 
colliding at the ALICE interaction point. The bunch intensity was
typically $7\times 10^7$ Pb ions per bunch. The luminosity varied within  
$0.5-8 \times 10^{23}$~cm$^{-2}$s$^{-1}$. 

The detector readout was activated by a minimum-bias interaction trigger based 
on signals measured in the forward scintillators (VZERO) and in the Silicon 
Pixel Detector (SPD), in coincidence with the LHC bunch-crossing signal. 
The VZERO counters are placed along the beam line 
at +3.3~m and -0.9~m from the interaction point. They cover the 
region $2.8<\eta<5.1$ (VZERO-A) and $-3.7<\eta<-1.7$ (VZERO-C) and record  
the amplitude and arrival time of signals produced by charged particles.
The inner and outer layers of the SPD cover the central pseudorapidity 
regions $|\eta|<2$ and $|\eta|<1.4$, respectively. 
The detector has a total of 9.8 million pixels read out by 1200 chips. 
Each chip provides a fast signal if at least one of its pixels is hit. 
The signals from the 1200 chips are combined in a programmable logic unit. 
The minimum-bias interaction trigger required at least two out of the
following three conditions: i) at least two pixel chips hit in the 
outer layer of the SPD, ii) a signal in VZERO-A, iii) a signal in
VZERO-C. 
More details of the trigger and run conditions are discussed 
in Ref.~\cite{Aamodt:2010pb}. 

For the present analysis we have used $1.6\times10^4$ events selected by requiring 
a primary vertex reconstructed within $\pm$12~cm of the nominal interaction point 
and applying a cut on the sum of the amplitudes measured in the VZERO detectors 
corresponding to the most central 5\% of the hadronic cross section. 
The \dens\ measured in this centrality class is 
\meandndeta=$1601 \pm 60$~(syst.) as published in Ref.~\cite{Collaboration:2010cz} 
where the centrality determination and the measurement of \dens\ are 
described in detail. 
The correlation analysis was performed using charged-particle tracks detected in 
the Inner Tracking System (ITS) and the Time Projection Chamber (TPC). 
The ITS extends over $3.9<r<43$~cm and contains, in addition to the two SPD 
layers described above, two layers of Silicon Drift 
Detectors and two layers of Silicon Strip Detectors, with $1.33\times 10^5$ and 
$2.6\times 10^6$ readout channels, respectively. 
The TPC is a cylindrical drift detector with two readout planes on the endcaps. 
The active volume covers $85<r<247$~cm and $-250<z<250$~cm in the radial 
and longitudinal directions, respectively. 
A high voltage membrane at $z=0$ divides the active volume into two halves and 
provides the electric drift field of 400~V/cm, resulting in a maximum drift time 
of 94~$\mu$s.
With the solenoidal magnetic field of $0.5$~T the momentum resolution 
for particles with \pt~$<1$\gevc\ is about 1\%.
Tracks at the edge of the acceptance were removed by restricting 
the analysis to the region $|\eta|<0.8$. 
Good track quality was ensured by requiring the tracks to have at least 
90 clusters in the TPC (out of a maximum of 159), to have at least two matching hits 
in the ITS (out of a maximum of 6), 
and to point back to the primary interaction vertex within 1~cm. 
In order to reduce the contamination of the pion sample by electrons and 
kaons, that would dilute the Bose-Einstein enhancement in the correlation 
function, we applied a cut on the specific ionization (\dedx) in the 
TPC gas. In central \pbpb\ collisions the \dedx\ resolution of the TPC is 
better than 7\%. 

\section{\label{sec:cor}Two-pion correlation functions}

The two-particle correlation function is defined as the ratio 
$C\left({\bf q}\right)=A\left({\bf q}\right)/B\left({\bf q}\right)$, 
where  $A\left({\bf q}\right)$ is the measured distribution of the 
difference ${\bf q}={\bf p}_2-{\bf p}_1$ between the three-momenta of the two 
particles ${\bf p}_1$ and ${\bf p}_2$, 
and $B\left({\bf q}\right)$ is the corresponding distribution formed by using 
pairs of particles where each particle comes from a different event 
(event mixing)~\cite{Kopylov:1974th}. 
Every event was mixed with five other events, and for each pair of events 
all pion candidates from one event were paired with all pion candidates 
from the other. 
The correlation functions were studied in bins of transverse momentum, 
defined as half the modulus of the vector sum of the two transverse momenta, 
\kt~=~$|{\bf p}_{T,1}+{\bf p}_{T,2}|/2$. 
The momentum difference is calculated in the longitudinally co-moving system 
(LCMS), where the longitudinal pair momentum vanishes,  
and is decomposed into (\Qout, \Qside, \Qlong), with the ``out'' axis pointing 
along the pair transverse momentum, the ``side'' axis perpendicular to it in 
the transverse plane, and the ``long'' axis along the beam 
(Bertsch--Pratt convention~\cite{Bertsch:1989vn,Pratt:1986cc}). 

Track splitting (incorrect reconstruction of a signal produced by one 
particle as two tracks) and track merging (reconstructing one track instead 
of two) generally lead to structures in the 
two-particle correlation functions if not properly treated. 
With the particular track selection used in this analysis, 
the track splitting effect is negligible and the track merging 
leads to a 20-30\% loss of track pairs with a distance 
of closest approach in the TPC of 1~cm or less.  
We have solved this problem by including in 
$A\left({\bf q}\right)$ and $B\left({\bf q}\right)$ only those 
track pairs that are separated by at least 1.2 cm in $r\Delta\phi$ 
or at least 2.4~cm in $z$ at a radius of 1.2~m. 
We have checked that with this selection one recovers the flat 
shape of the correlation function in Monte Carlo simulations 
that do not include Bose--Einstein enhancement. 

\begin{figure}[t]
\centerline{\includegraphics[width=0.8\textwidth]{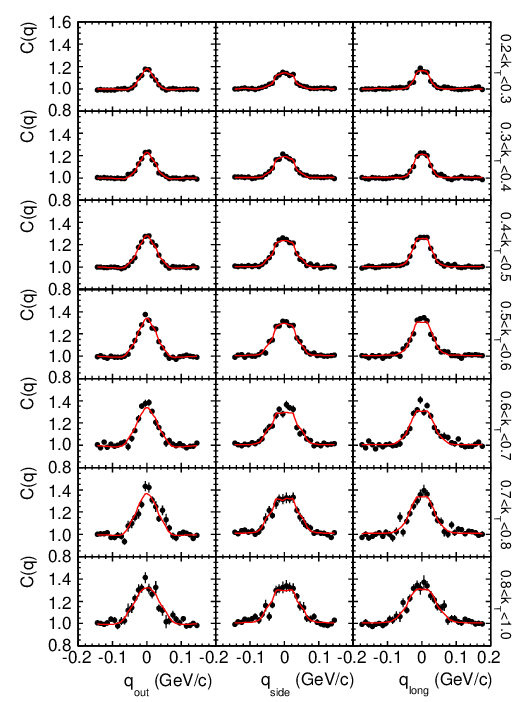}}
\caption{\label{fig:cor}
Projections of the three-dimensional \pim\pim\ correlation function (points) 
and of the respective fits (lines) for seven \kt\ intervals. 
When projecting on one axis the other two components were required to be 
within (-0.03,0.03)\gevc. 
The \kt\ range is indicated on the right-hand side axis in\gevc.}
\end{figure}

Projections of three-dimensional \pim\pim\ correlation functions  
$C$(\Qout, \Qside, \Qlong) for seven \kt\ bins from 0.2 to 1.0\gevc\
are shown in Fig.~\ref{fig:cor}. 
The correlation functions for positive pion pairs look similar. 
The Bose--Einstein enhancement peak is manifest at low $q=|{\bf q}|$. 
The peak width increases when going from low to high transverse momenta. 
The three-dimensional correlation functions were fitted by an 
expression~\cite{Sinyukov:1998fc} accounting for the Bose--Einstein 
enhancement and for the Coulomb interaction between the two particles: 
\begin{eqnarray}
\label{eq:1dbowler-sinyukov}
\nonumber C({\rm \bf q})&=& \mathcal{N} \left[ (1-\lambda)+\lambda K(q_{\rm inv}) (1 + G({\rm \bf q}) ) \right] ,\\
\nonumber G({\rm \bf q}) &=& \exp(-(\rout^2 \qout^2 + \rside^2 \qside^2 + \rlong^2 \qlong^2 + \\
&&+2|\rol| \rol \qout\qlong)),
\end{eqnarray}
with $\lambda$ describing the correlation strength, and \Rout, 
\Rside, and \Rlong\ being the Gaussian HBT radii. 
The parameter \Rol, that quantifies the cross term between 
\Qout\ and \Qlong, was found to be consistent with zero, as expected for 
measurements at midrapidity in a symmetric system. 
This term was therefore set equal to zero in the final fits. 
The factor $K(q_{\rm inv})$ is the squared Coulomb wave function 
averaged over a spherical source~\cite{Pratt:1986ev} of size equal 
to the mean of \Rout, \Rside, and \Rlong; its argument $q_{\rm inv}$, 
for pairs of identical pions, is equal to $q$ calculated in the pair rest frame. 
The Coulomb effect is taken to be attenuated by the same factor 
$\lambda$ as the Bose--Einstein peak. 
The fit function is shown as a solid line in Fig.~\ref{fig:cor}. 

The obtained radii have been corrected for the finite momentum 
resolution that smears out the correlation peak. 
The effect was studied by applying weights to pairs of tracks in simulated 
events so as to produce the correlation function expected for a given 
set of the HBT radii. The weights were calculated using the original 
Monte Carlo momenta. The reconstructed radii were found to differ from the 
input ones by up to 4\%, depending on the radius and \kt. 
The corresponding correction was applied to the experimental HBT radii. 

\section{\label{sec:syst}Systematic uncertainties}
The systematic uncertainties on the HBT parameters were estimated by comparing 
the results obtained by varying the analysis procedure. 
Not requiring the ITS hits in the tracking leads to a variation of the transverse 
and longitudinal radii of up to 3\% and 8\%, respectively. 
Variation of the pion identification criteria within a reasonable range 
introduces radius variations of up to 5\%. 
Changing the fit range in $q$ from 0-0.3\gevc\ to 0-0.5\gevc\ results in a reduction 
of all three radii by about 3\%. 
Increasing the two-track separation cut by 50\% results in a change 
of the radii by up to 3\%. 
Generating the denominator of the correlation function by rotating one of the 
two tracks by 180\deg\ rather than by event mixing results in an increase 
of 6\% for \Rside\ at low \kt\ and up to 4\% for \Rout\ and \Rlong.  
The systematic error connected with the Coulomb correction was evaluated by 
modifying the source radius used for the correction by $\pm$2~fm. This was 
found to affect mostly \Rout\ which changed by up to 4\%. 
The correction for the momentum resolution is about 4\%. The corresponding 
uncertainty on the final radii, tested by modifying the momentum 
resolution by 20\%, is negligible. 
Finally, a study performed with an independent analysis code, including a 
different pair selection criterion (accepting only those 50\% of the pairs 
for which the $r\Delta\phi$ separation between the two tracks increases 
with the radius, and requiring that the separation is at least 2~cm 
at the entrance to the TPC), 
yields transverse radii and \Rlong\ that differ by up to 5\% and 8\%, 
respectively. 
The total systematic errors are estimated by adding up the mentioned 
contributions in quadrature and are largest (9-10\%) for 
the transverse radii in the lowest \kt\ bin and for \Rlong\ above 0.65\gevc. 

\section{\label{sec:ktdep}Transverse momentum dependence of the radii}
\begin{figure}[t]
\centerline{\includegraphics[width=0.65\textwidth]{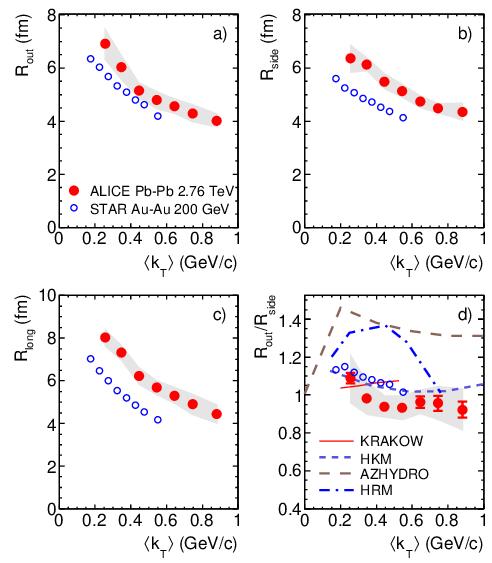}}
\caption{\label{fig:ktdep}
Pion HBT radii for the 5\% most central \pbpb\ collisions 
at \ene, as function of \meankt\ (red filled dots). The shaded bands represent 
the systematic errors. For comparison, parameters for \auau\ collisions 
at \sqrts~=~200~GeV~\cite{Adams:2004yc} are shown as blue open circles. 
(The combined, statistical and systematic, errors on these measurements  
are below 4\%.)
The lines show model predictions (see text). }
\end{figure}

The HBT radii extracted from the fit to the two-pion correlation functions 
and corrected for the momentum resolution 
as described in the previous section 
are shown as a function of \meankt\ in Table~\ref{tab:ktdep} and in Fig.~\ref{fig:ktdep}. 
\begin{table}[b]
\caption{\label{tab:ktdep}
Pion HBT radii for the 5\% most central \pbpb\ collisions at \ene, as function 
of \meankt. The first error is statistical and the second is systematic.} 
\begin{center}
\begin{tabular}{cccc}
\hline
\meankt (GeV/$c$) &     \Rout~(fm)        &        \Rside~(fm)           &        \Rlong~(fm)          \\
\hline
0.26 &  6.92 $\pm$ 0.12 $\pm$ 0.61 &  6.36 $\pm$ 0.12 $\pm$ 0.54 &  8.03 $\pm$ 0.15 $\pm$ 0.42\\
0.35 &  6.03 $\pm$ 0.08 $\pm$ 0.48 &  6.13 $\pm$ 0.09 $\pm$ 0.26 &  7.31 $\pm$ 0.10 $\pm$ 0.39\\
0.44 &  5.15 $\pm$ 0.07 $\pm$ 0.30 &  5.49 $\pm$ 0.08 $\pm$ 0.30 &  6.23 $\pm$ 0.09 $\pm$ 0.41\\
0.54 &  4.79 $\pm$ 0.08 $\pm$ 0.34 &  5.14 $\pm$ 0.09 $\pm$ 0.26 &  5.67 $\pm$ 0.10 $\pm$ 0.35\\
0.64 &  4.56 $\pm$ 0.10 $\pm$ 0.29 &  4.73 $\pm$ 0.11 $\pm$ 0.25 &  5.30 $\pm$ 0.12 $\pm$ 0.40\\
0.75 &  4.29 $\pm$ 0.12 $\pm$ 0.34 &  4.48 $\pm$ 0.13 $\pm$ 0.20 &  4.90 $\pm$ 0.15 $\pm$ 0.50\\
0.88 &  4.02 $\pm$ 0.14 $\pm$ 0.26 &  4.35 $\pm$ 0.14 $\pm$ 0.34 &  4.43 $\pm$ 0.15 $\pm$ 0.45\\
\hline
\end{tabular}
\end{center}
\end{table}
The fit parameters for positive and negative pion pairs agree within statistical 
errors and therefore the averages are presented here. 
The HBT radii for the 5\% most central \pbpb\ collisions at \ene\ 
are found to be significantly (10-35\%) larger than those measured by STAR in 
central \auau\ collisions at \sqrts~=~200~GeV~\cite{Adams:2004yc}. 
The increase is beyond systematic errors and is similarly strong for 
\Rside\ and \Rlong. 
As also observed in heavy-ion collision experiments at lower 
energies~\cite{Lisa:2005dd}, the HBT radii show a decreasing trend with 
increasing \kt. 
This is a characteristic feature of expanding particle sources since the 
HBT radii describe the homogeneity length rather than the overall size of 
the particle-emitting system~\cite{Makhlin:1987gm,Sinyukov:1989xz,Akkelin:1995gh,Sinyukov:1994vg}. 
The homogeneity length is defined as the size of the region that contributes 
to the pion spectrum at a particular three-momentum {\bf p}. 
The \Rout\ radius is comparable with \Rside\ and the
\kt\ dependence of the ratio \Rout/\Rside\ is flat within the systematic 
errors. \Rlong\ is seen to be somewhat larger than \Rout\ and 
\Rside\ and to decrease slightly faster with increasing \kt.

The extracted $\lambda$-parameter is found to range from 0.5 to 0.7 
and increases slightly with \kt. Somewhat lower values but 
a similar \kt\ dependence were observed in \auau\ collisions at 
RHIC~\cite{Adams:2004yc}. 

\section{\label{sec:enedep}Beam energy dependence of the radii}
\begin{figure}[t]
\centerline{\includegraphics[width=0.52\textwidth]{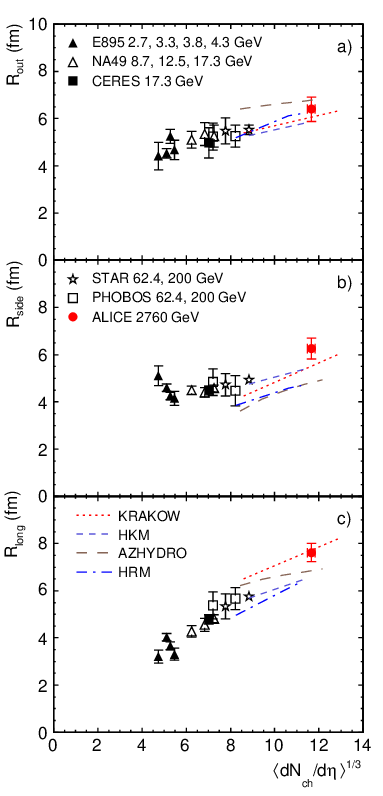}}
\caption{
Pion HBT radii at \kt~=~0.3\gevc\ for the 5\% most central \pbpb\ at \ene\ 
(red filled dot) and the radii obtained for central gold and lead collisions 
at lower energies at the 
AGS~\cite{Lisa:2000no}, 
SPS~\cite{
Alt:2007uj,         
Afanasiev:2002mx,   
Adamova:2002wi},    
and RHIC~\cite{
Abelev:2009tp,      
Back:2004ug,        
Back:2005hs,        
Back:2002wb,        
Adams:2004yc,       
Abelev:2008ez}.     
Model predictions are shown as lines. 
\label{fig:nchdep}}
\end{figure}

In Fig.~\ref{fig:nchdep}, we compare the three radii at \meankt~=~0.3\gevc\ 
with experimental results at lower energies. 
The values of the radii at this \kt\ were obtained by parabolic interpolation.  
Following the established practice~\cite{Lisa:2005dd} we plot the radii as 
functions of \meandndeta$^{1/3}$.  
In this representation the comparison is not affected by slight differences 
between the mass numbers of the colliding nuclei and between centralities. 
For E895 and NA49, \dndeta\ has been approximated 
using the published rapidity densities. 
The reference frame dependence of \dndeta\ is neglected. 
The errors on the E895 points are statistical only. 
For the other experiments the error 
bars represent the statistical and systematic uncertainties 
added in quadrature. 
For the ALICE point the error is dominated by the systematic uncertainties. 

The ALICE measurement significantly extends the range of the existing 
world systematics of HBT radii. 
The trend of \Rlong\ growing approximately linearly with the cube 
root of the \dens, established at lower energies, continues at the LHC 
(Fig.~\ref{fig:nchdep}-c). 
The situation is similar with \Rout\ (Fig.~\ref{fig:nchdep}-a) which 
also grows with energy albeit slower than \Rlong. 
For \Rside, that is most directly related to the transverse size of the 
pion source and is less affected by experimental uncertainties, 
an increase is observed beyond systematic errors (Fig.~\ref{fig:nchdep}-b). 
At lower energies a rather flat behavior with a shallow minimum between 
AGS and SPS energies was observed and interpreted as due to the transition 
from baryon to meson dominance at freeze-out~\cite{Adamova:2002ff}. 
An increase of \Rside\ at high energy is consistent with that interpretation.

Available model predictions are compared to the experimental data in 
Figs.~\ref{fig:ktdep}-d and \ref{fig:nchdep}. 
Calculations from three models incorporating a hydrodynamic approach, 
AZHYDRO~\cite{Frodermann:2007ab}, 
KRAKOW~\cite{Bozek:2010wt,Chojnacki:2007rq}, 
and HKM~\cite{Karpenko:2009wf,Abreu:2007kv}, 
and from the hadronic-kinematics-based model HRM~\cite{Humanic:2008nt,Humanic:2010su} 
are shown. 
An in-depth discussion is beyond the scope of this Letter but we notice that, 
while the increase of the radii between RHIC and the LHC is roughly reproduced by 
all four calculations, only two of them (KRAKOW and HKM) are able to describe 
the experimental \Rout/\Rside\ ratio. 

\begin{figure}
\centerline{\includegraphics[width=0.7\textwidth]{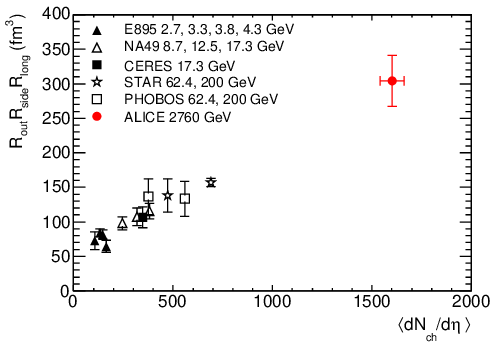}}
\caption{
Product of the three pion HBT radii at \kt~=~0.3\gevc. 
The ALICE result (red filled dot) is compared to those obtained 
for central gold and lead collisions at lower energies 
at the 
AGS~\cite{Lisa:2000no}, 
SPS~\cite{
Alt:2007uj,         
Afanasiev:2002mx,   
Adamova:2002wi},    
and RHIC~\cite{
Abelev:2009tp,      
Back:2004ug,        
Back:2005hs,        
Back:2002wb,        
Adams:2004yc,       
Abelev:2008ez}.     
\label{fig:product}}
\end{figure}

The systematics of the product of the three radii is shown in 
Fig.~\ref{fig:product}. 
The product of the radii, which is connected to the volume of the 
homogeneity region, shows 
a linear dependence on the \dens\ and is 
two times larger at the LHC than at RHIC. 

\begin{figure}
\centerline{\includegraphics[width=0.7\textwidth]{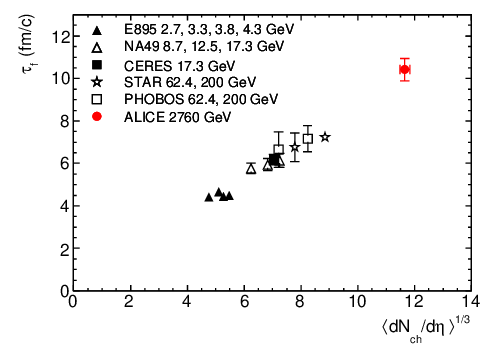}}
\caption{
The decoupling time extracted from \Rlong(\kt). 
The ALICE result (red filled dot) is compared to those obtained 
for central gold and lead collisions at lower energies 
at the 
AGS~\cite{Lisa:2000no}, 
SPS~\cite{
Alt:2007uj,         
Afanasiev:2002mx,   
Adamova:2002wi},    
and RHIC~\cite{
Abelev:2009tp,      
Back:2004ug,        
Back:2005hs,        
Back:2002wb,        
Adams:2004yc,       
Abelev:2008ez}.     
\label{fig:tau}}
\end{figure}

Within hydrodynamic scenarios, 
the decoupling time for hadrons at midrapidity can be estimated in the following way. 
The size of the homogeneity region is inversely proportional to the velocity 
gradient of the expanding system. 
The longitudinal velocity gradient 
in a high energy nuclear collision decreases with time as 1/$\tau$~\cite{Bjorken:1982qr}. 
Therefore, the magnitude of \Rlong\ is proportional to the total duration of 
the longitudinal expansion, i.e. to the decoupling time of the system~\cite{Makhlin:1987gm}. 
Quantitatively, the decoupling time \tauf\ can be obtained by fitting \Rlong\ with 
\begin{equation}
\label{eq:rlongkt}
\rlong{}^2(k_T) = \frac{\tau_f^2 T}{m_T} \frac{K_2(m_T/T)}{K_1(m_T/T)} \ , \ \ m_T=\sqrt{m_{\pi}^2+k_T^2}, 
\end{equation}
where $m_\pi$ is the pion mass, $T$ the kinetic freeze-out temperature taken to be 0.12 GeV, and 
$K_1$ and $K_2$ are the integer order modified Bessel functions~\cite{Makhlin:1987gm,Herrmann:1994rr}. 
The decoupling time extracted from this fit to the ALICE radii 
and to the values published at lower energies are shown in Figure~\ref{fig:tau}. 
As can be seen, \tauf\  
scales with the cube root of \dens\  and reaches 10--11~fm/$c$ 
in central \pbpb\ collisions at \ene.
It should be kept in mind that 
while Eq.~(\ref{eq:rlongkt}) captures basic features of a longitudinally expanding 
particle-emitting system, in the presence of transverse expansion 
and a finite chemical potential of pions it may underestimate the actual decoupling time 
by about 25\%~\cite{Sinyukov:1996ww}. 
An  uncertainty is connected to the value of the kinetic freeze-out temperature 
used in the fit $T=0.12$~GeV. 
Setting $T$ to 0.1~GeV~\cite{Chapman:1996ec,Alt:2007uj,Adams:2004yc,Retiere:2003kf} 
and 0.14~GeV~\cite{Broniowski:2008vp} leads to a $\tau_f$ value that is 13\% higher 
and 10\% lower, respectively. 

\section{\label{sec:sum}Summary}

We have presented the first analysis of the two-pion correlation functions 
in \pbpb\ collisions at \ene\ at the LHC. The pion source radii 
obtained from this measurement exceed those measured at RHIC by 10-35\%. 
The increase is beyond systematic errors and is present for both the 
longitudinal and transverse radii. 
The homogeneity volume is found to be larger by a factor of two. 
The decoupling time for midrapidity pions exceeds 10~fm/$c$ which is 
40\% larger than at RHIC. 
These results, taken together with those obtained from the study of 
multiplicity~\cite{Aamodt:2010pb,Collaboration:2010cz} and the azimuthal 
anisotropy~\cite{Aamodt:2010pa}, 
indicate that the fireball formed in nuclear collisions at the LHC is hotter, 
lives longer, and expands to a larger size at freeze-out as compared to lower 
energies.

\section*{Acknowledgements}
\input{acknowledgements.tex}

\input{myfemto.bbl}
\newpage

\appendix
\section{The ALICE Collaboration}
\label{app:collab}
\input{authors-cernphprep.tex} 
\end{document}

%% file: acknowledgements.tex
The ALICE Collaboration would like to thank all its engineers and technicians for their invaluable contributions to the construction of the experiment and the CERN accelerator teams for the outstanding performance of the LHC complex.
The ALICE Collaboration acknowledges the following funding agencies for their support in building and
running the ALICE detector:
Calouste Gulbenkian Foundation from Lisbon and Swiss Fonds Kidagan, Armenia;
Conselho Nacional de Desenvolvimento Cient\'{\i}fico e Tecnol\'{o}gico (CNPq), Financiadora de Estudos e Projetos (FINEP),
Funda\c{c}\~{a}o de Amparo \`{a} Pesquisa do Estado de S\~{a}o Paulo (FAPESP);
National Natural Science Foundation of China (NSFC), the Chinese Ministry of Education (CMOE)
and the Ministry of Science and Technology of China (MSTC);
Ministry of Education and Youth of the Czech Republic;
Danish Natural Science Research Council, the Carlsberg Foundation and the Danish National Research Foundation;
The European Research Council under the European Community's Seventh Framework Programme;
Helsinki Institute of Physics and the Academy of Finland;
French CNRS-IN2P3, the `Region Pays de Loire', `Region Alsace', `Region Auvergne' and CEA, France;
German BMBF and the Helmholtz Association;
ExtreMe Matter Institute EMMI, Germany;
Greek Ministry of Research and Technology;
Hungarian OTKA and National Office for Research and Technology (NKTH);
Department of Atomic Energy and Department of Science and Technology of the Government of India;
Istituto Nazionale di Fisica Nucleare (INFN) of Italy;
MEXT Grant-in-Aid for Specially Promoted Research, Ja\-pan;
Joint Institute for Nuclear Research, Dubna;
 %
National Research Foundation of Korea (NRF);
CONACYT, DGAPA, M\'{e}xico, ALFA-EC and the HELEN Program (High-Energy physics Latin-American--European Network);
Stichting voor Fundamenteel Onderzoek der Materie (FOM) and the Nederlandse Organisatie voor Wetenschappelijk Onderzoek (NWO), Netherlands;
Research Council of Norway (NFR);
Polish Ministry of Science and Higher Education;
National Authority for Scientific Research - NASR (Autoritatea Na\c{t}ional\u{a} pentru Cercetare \c{S}tiin\c{t}ific\u{a} - ANCS);
Federal Agency of Science of the Ministry of Education and Science of Russian Federation, International Science and
Technology Center, Russian Academy of Sciences, Russian Federal Agency of Atomic Energy, Russian Federal Agency for Science and Innovations and CERN-INTAS;
Ministry of Education of Slovakia;
CIEMAT, EELA, Ministerio de Educaci\'{o}n y Ciencia of Spain, Xunta de Galicia (Conseller\'{\i}a de Educaci\'{o}n),
CEA\-DEN, Cubaenerg\'{\i}a, Cuba, and IAEA (International Atomic Energy Agency);
The Ministry of Science and Technology and the National Research Foundation (NRF), South Africa;
Swedish Research Council (VR) and Knut $\&$ Alice Wallenberg Foundation (KAW);
Ukraine Ministry of Education and Science;
United Kingdom Science and Technology Facilities Council (STFC);
The United States Department of Energy, the United States National
Science Foundation, the State of Texas, and the State of Ohio.

%% file: authors-cernphprep.tex
%
\begingroup
\small
\begin{flushleft}
K.~Aamodt\Irefn{0}\And
A.~Abrahantes~Quintana\Irefn{1}\And
D.~Adamov\'{a}\Irefn{2}\And
A.M.~Adare\Irefn{3}\And
M.M.~Aggarwal\Irefn{4}\And
G.~Aglieri~Rinella\Irefn{5}\And
A.G.~Agocs\Irefn{6}\And
S.~Aguilar~Salazar\Irefn{7}\And
Z.~Ahammed\Irefn{8}\And
N.~Ahmad\Irefn{9}
% ...
\renewcommand\labelenumi{\textsuperscript{\theenumi}~}
\section*{Affiliation notes}
\renewcommand\theenumi{\roman{enumi}}
\begin{Authlist}
\item \Adef{0}Also at Laboratoire de Physique Corpusculaire (LPC), Clermont Universit\'{e}, Universit\'{e} Blaise Pascal, CNRS--IN2P3, Clermont-Ferrand, France
\item \Adef{1}Now at Centro Fermi -- Centro Studi e Ricerche e Museo Storico della Fisica ``Enrico Fermi'', Rome, Italy
\item \Adef{2}Now at European Organization for Nuclear Research (CERN), Geneva, Switzerland
\item \Adef{3}Also at European Organization for Nuclear Research (CERN), Geneva, Switzerland
\item \Adef{4}Now at Physikalisches Institut, Ruprecht-Karls-Universit\"{a}t Heidelberg, Heidelberg, Germany
\item \Adef{5}Now at Sezione INFN, Turin, Italy
\item \Adef{6}Now at University of Houston, Houston, Texas, United States
\item \Adef{7}Also at Dipartimento di Fisica Sperimentale dell'Universit\`{a} and Sezione INFN, Turin, Italy
\item \Adef{8}Also at  Dipartimento di Fisica dell'Universit\`{a}, Udine, Italy 
\item \Adef{9}Now at SUBATECH, Ecole des Mines de Nantes, Universit\'{e} de Nantes, CNRS-IN2P3, Nantes, France
\item \Adef{10}Now at Centro de Investigaci\'{o}n y de Estudios Avanzados (CINVESTAV), Mexico City and M\'{e}rida, Mexico
\item \Adef{11}Now at Benem\'{e}rita Universidad Aut\'{o}noma de Puebla, Puebla, Mexico
\item \Adef{12}Now at Laboratoire de Physique Subatomique et de Cosmologie (LPSC), Universit\'{e} Joseph Fourier, CNRS-IN2P3, Institut Polytechnique de Grenoble, Grenoble, France
\item \Adef{13}Now at Institut Pluridisciplinaire Hubert Curien (IPHC), Universit\'{e} de Strasbourg, CNRS-IN2P3, Strasbourg, France
\item \Adef{14}Now at Sezione INFN, Padova, Italy
\item \Adef{15} Deceased 
\item \Adef{16}Also at Division of Experimental High Energy Physics, University of Lund, Lund, Sweden
\item \Adef{17}Also at  University of Technology and Austrian Academy of Sciences, Vienna, Austria 
\item \Adef{18}Now at Oak Ridge National Laboratory, Oak Ridge, Tennessee, United States
\item \Adef{19}Also at Wayne State University, Detroit, Michigan, United States
\item \Adef{20}Also at Frankfurt Institute for Advanced Studies, Johann Wolfgang Goethe-Universit\"{a}t Frankfurt, Frankfurt, Germany
\item \Adef{21}Now at Frankfurt Institute for Advanced Studies, Johann Wolfgang Goethe-Universit\"{a}t Frankfurt, Frankfurt, Germany
\item \Adef{22}Now at Research Division and ExtreMe Matter Institute EMMI, GSI Helmholtzzentrum f\"ur Schwerionenforschung, Darmstadt, Germany
\item \Adef{23}Also at Fachhochschule K\"{o}ln, K\"{o}ln, Germany
\item \Adef{24}Also at Institute of Experimental Physics, Slovak Academy of Sciences, Ko\v{s}ice, Slovakia
\item \Adef{25}Now at Instituto de Ciencias Nucleares, Universidad Nacional Aut\'{o}noma de M\'{e}xico, Mexico City, Mexico
\item \Adef{26}Also at Laboratoire de Physique Subatomique et de Cosmologie (LPSC), Universit\'{e} Joseph Fourier, CNRS-IN2P3, Institut Polytechnique de Grenoble, Grenoble, France
\item \Adef{27}Also at  "Vin\v{c}a" Institute of Nuclear Sciences, Belgrade, Serbia 
\item \Adef{28}Also at Dipartimento di Fisica `E.R.~Caianiello' dell'Universit\`{a} and Gruppo Collegato INFN, Salerno, Italy
\item \Adef{29}Also at Instituto de Ciencias Nucleares, Universidad Nacional Aut\'{o}noma de M\'{e}xico, Mexico City, Mexico
\item \Adef{30}Also at University of Houston, Houston, Texas, United States
\item \Adef{31}Also at Department of Physics, University of Oslo, Oslo, Norway
\item \Adef{32}Now at Department of Physics, University of Oslo, Oslo, Norway
\item \Adef{33}Also at Eberhard Karls Universit\"{a}t T\"{u}bingen, T\"{u}bingen, Germany
\item \Adef{34}Also at Dipartimento Interateneo di Fisica `M.~Merlin' and Sezione INFN, Bari, Italy
\item \Adef{35}Now at Nikhef, National Institute for Subatomic Physics and Institute for Subatomic Physics of Utrecht University, Utrecht, Netherlands
\item \Adef{36}Also at Hua-Zhong Normal University, Wuhan, China
\item \Adef{37}Also at Centro Fermi -- Centro Studi e Ricerche e Museo Storico della Fisica ``Enrico Fermi'', Rome, Italy
\item \Adef{38}Also at M.V.~Lomonosov Moscow State University, D.V.~Skobeltsyn Institute of Nuclear Physics, Moscow, Russia
\end{Authlist}
\section*{Collaboration Institutes}
\renewcommand\theenumi{\arabic{enumi}~}
\begin{Authlist}
\item \Idef{0}Department of Physics and Technology, University of Bergen, Bergen, Norway
\item \Idef{1}Centro de Aplicaciones Tecnol\'{o}gicas y Desarrollo Nuclear (CEADEN), Havana, Cuba
\item \Idef{2}Nuclear Physics Institute, Academy of Sciences of the Czech Republic, \v{R}e\v{z} u Prahy, Czech Republic
\item \Idef{3}Yale University, New Haven, Connecticut, United States
\item \Idef{4}Physics Department, Panjab University, Chandigarh, India
\item \Idef{5}European Organization for Nuclear Research (CERN), Geneva, Switzerland
\item \Idef{6}KFKI Research Institute for Particle and Nuclear Physics, Hungarian Academy of Sciences, Budapest, Hungary
\item \Idef{7}Instituto de F\'{\i}sica, Universidad Nacional Aut\'{o}noma de M\'{e}xico, Mexico City, Mexico
\item \Idef{8}Variable Energy Cyclotron Centre, Kolkata, India
\item \Idef{9}Department of Physics, Aligarh Muslim University, Aligarh, India
\item \Idef{10}Gangneung-Wonju National University, Gangneung, South Korea
\item \Idef{11}Institute for Theoretical and Experimental Physics, Moscow, Russia
\item \Idef{12}Russian Research Centre Kurchatov Institute, Moscow, Russia
\item \Idef{13}Sezione INFN, Turin, Italy
\item \Idef{14}Dipartimento di Fisica dell'Universit\`{a} and Sezione INFN, Bologna, Italy
\item \Idef{15}Bogolyubov Institute for Theoretical Physics, Kiev, Ukraine
\item \Idef{16}Frankfurt Institute for Advanced Studies, Johann Wolfgang Goethe-Universit\"{a}t Frankfurt, Frankfurt, Germany
\item \Idef{17}Dipartimento Interateneo di Fisica `M.~Merlin' and Sezione INFN, Bari, Italy
\item \Idef{18}Research Division and ExtreMe Matter Institute EMMI, GSI Helmholtzzentrum f\"ur Schwerionenforschung, Darmstadt, Germany
\item \Idef{19}V.~Fock Institute for Physics, St. Petersburg State University, St. Petersburg, Russia
\item \Idef{20}National Institute for Physics and Nuclear Engineering, Bucharest, Romania
\item \Idef{21}Kirchhoff-Institut f\"{u}r Physik, Ruprecht-Karls-Universit\"{a}t Heidelberg, Heidelberg, Germany
\item \Idef{22}Department of Physics, Ohio State University, Columbus, Ohio, United States
\item \Idef{23}Rudjer Bo\v{s}kovi\'{c} Institute, Zagreb, Croatia
\item \Idef{24}Sezione INFN, Padova, Italy
\item \Idef{25}Sezione INFN, Bologna, Italy
\item \Idef{26}SUBATECH, Ecole des Mines de Nantes, Universit\'{e} de Nantes, CNRS-IN2P3, Nantes, France
\item \Idef{27}Institut f\"{u}r Kernphysik, Johann Wolfgang Goethe-Universit\"{a}t Frankfurt, Frankfurt, Germany
\item \Idef{28}Laboratoire de Physique Subatomique et de Cosmologie (LPSC), Universit\'{e} Joseph Fourier, CNRS-IN2P3, Institut Polytechnique de Grenoble, Grenoble, France
\item \Idef{29}Departamento de F\'{\i}sica de Part\'{\i}culas and IGFAE, Universidad de Santiago de Compostela, Santiago de Compostela, Spain
\item \Idef{30}Oak Ridge National Laboratory, Oak Ridge, Tennessee, United States
\item \Idef{31}Helsinki Institute of Physics (HIP) and University of Jyv\"{a}skyl\"{a}, Jyv\"{a}skyl\"{a}, Finland
\item \Idef{32}Sezione INFN, Catania, Italy
\item \Idef{33}Dipartimento di Fisica Sperimentale dell'Universit\`{a} and Sezione INFN, Turin, Italy
\item \Idef{34}Centro Fermi -- Centro Studi e Ricerche e Museo Storico della Fisica ``Enrico Fermi'', Rome, Italy
\item \Idef{35}Commissariat \`{a} l'Energie Atomique, IRFU, Saclay, France
\item \Idef{36}Laboratoire de Physique Corpusculaire (LPC), Clermont Universit\'{e}, Universit\'{e} Blaise Pascal, CNRS--IN2P3, Clermont-Ferrand, France
\item \Idef{37}Institute of Experimental Physics, Slovak Academy of Sciences, Ko\v{s}ice, Slovakia
\item \Idef{38}Dipartimento di Fisica e Astronomia dell'Universit\`{a} and Sezione INFN, Catania, Italy
\item \Idef{39}School of Physics and Astronomy, University of Birmingham, Birmingham, United Kingdom
\item \Idef{40}The Henryk Niewodniczanski Institute of Nuclear Physics, Polish Academy of Sciences, Cracow, Poland
\item \Idef{41}Institut f\"{u}r Kernphysik, Westf\"{a}lische Wilhelms-Universit\"{a}t M\"{u}nster, M\"{u}nster, Germany
\item \Idef{42}Joint Institute for Nuclear Research (JINR), Dubna, Russia
\item \Idef{43}Niels Bohr Institute, University of Copenhagen, Copenhagen, Denmark
\item \Idef{44}Institut Pluridisciplinaire Hubert Curien (IPHC), Universit\'{e} de Strasbourg, CNRS-IN2P3, Strasbourg, France
\item \Idef{45}Wayne State University, Detroit, Michigan, United States
\item \Idef{46}Petersburg Nuclear Physics Institute, Gatchina, Russia
\item \Idef{47}Physics Department, University of Jammu, Jammu, India
\item \Idef{48}Laboratori Nazionali di Frascati, INFN, Frascati, Italy
\item \Idef{49}Dipartimento di Fisica dell'Universit\`{a} and Sezione INFN, Padova, Italy
\item \Idef{50}Faculty of Nuclear Sciences and Physical Engineering, Czech Technical University in Prague, Prague, Czech Republic
\item \Idef{51}Nikhef, National Institute for Subatomic Physics, Amsterdam, Netherlands
\item \Idef{52}Centro de Investigaciones Energ\'{e}ticas Medioambientales y Tecnol\'{o}gicas (CIEMAT), Madrid, Spain
\item \Idef{53}University of Houston, Houston, Texas, United States
\item \Idef{54}Moscow Engineering Physics Institute, Moscow, Russia
\item \Idef{55}Institute for High Energy Physics, Protvino, Russia
\item \Idef{56}Faculty of Science, P.J.~\v{S}af\'{a}rik University, Ko\v{s}ice, Slovakia
\item \Idef{57}Saha Institute of Nuclear Physics, Kolkata, India
\item \Idef{58}Institut de Physique Nucl\'{e}aire d'Orsay (IPNO), Universit\'{e} Paris-Sud, CNRS-IN2P3, Orsay, France
\item \Idef{59}Department of Physics, University of Oslo, Oslo, Norway
\item \Idef{60}Dipartimento di Fisica dell'Universit\`{a} and Sezione INFN, Trieste, Italy
\item \Idef{61}Faculty of Mathematics, Physics and Informatics, Comenius University, Bratislava, Slovakia
\item \Idef{62}Russian Federal Nuclear Center (VNIIEF), Sarov, Russia
\item \Idef{63}Physikalisches Institut, Ruprecht-Karls-Universit\"{a}t Heidelberg, Heidelberg, Germany
\item \Idef{64}Physics Department, University of Cape Town, iThemba LABS, Cape Town, South Africa
\item \Idef{65}Hua-Zhong Normal University, Wuhan, China
\item \Idef{66}Secci\'{o}n F\'{\i}sica, Departamento de Ciencias, Pontificia Universidad Cat\'{o}lica del Per\'{u}, Lima, Peru
\item \Idef{67}Physics Department, Creighton University, Omaha, Nebraska, United States
\item \Idef{68}Universit\'{e} de Lyon, Universit\'{e} Lyon 1, CNRS/IN2P3, IPN-Lyon, Villeurbanne, France
\item \Idef{69}Universidade Estadual de Campinas (UNICAMP), Campinas, Brazil
\item \Idef{70}Nikhef, National Institute for Subatomic Physics and Institute for Subatomic Physics of Utrecht University, Utrecht, Netherlands
\item \Idef{71}Division of Experimental High Energy Physics, University of Lund, Lund, Sweden
\item \Idef{72}University of Tsukuba, Tsukuba, Japan
\item \Idef{73}Sezione INFN, Cagliari, Italy
\item \Idef{74}Centro de Investigaci\'{o}n y de Estudios Avanzados (CINVESTAV), Mexico City and M\'{e}rida, Mexico
\item \Idef{75}Benem\'{e}rita Universidad Aut\'{o}noma de Puebla, Puebla, Mexico
\item \Idef{76}Dipartimento di Scienze e Tecnologie Avanzate dell'Universit\`{a} del Piemonte Orientale and Gruppo Collegato INFN, Alessandria, Italy
\item \Idef{77}Instituto de Ciencias Nucleares, Universidad Nacional Aut\'{o}noma de M\'{e}xico, Mexico City, Mexico
\item \Idef{78}Laboratori Nazionali di Legnaro, INFN, Legnaro, Italy
\item \Idef{79}Institute of Space Sciences (ISS), Bucharest, Romania
\item \Idef{80}Institute of Physics, Bhubaneswar, India
\item \Idef{81}Universidade de S\~{a}o Paulo (USP), S\~{a}o Paulo, Brazil
\item \Idef{82}Dipartimento di Fisica `E.R.~Caianiello' dell'Universit\`{a} and Gruppo Collegato INFN, Salerno, Italy
\item \Idef{83}Sezione INFN, Bari, Italy
\item \Idef{84}Dipartimento di Fisica dell'Universit\`{a} and Sezione INFN, Cagliari, Italy
\item \Idef{85}Soltan Institute for Nuclear Studies, Warsaw, Poland
\item \Idef{86}Sezione INFN, Rome, Italy
\item \Idef{87}Faculty of Engineering, Bergen University College, Bergen, Norway
\item \Idef{88}Institute for Nuclear Research, Academy of Sciences, Moscow, Russia
\item \Idef{89}Sezione INFN, Trieste, Italy
\item \Idef{90}Physics Department, University of Athens, Athens, Greece
\item \Idef{91}Warsaw University of Technology, Warsaw, Poland
\item \Idef{92}Universidad Aut\'{o}noma de Sinaloa, Culiac\'{a}n, Mexico
\item \Idef{93}Technical University of Split FESB, Split, Croatia
\item \Idef{94}Yerevan Physics Institute, Yerevan, Armenia
\item \Idef{95}University of Tokyo, Tokyo, Japan
\item \Idef{96}Department of Physics, Sejong University, Seoul, South Korea
\item \Idef{97}Lawrence Berkeley National Laboratory, Berkeley, California, United States
\item \Idef{98}Indian Institute of Technology, Mumbai, India
\item \Idef{99}Institut f\"{u}r Kernphysik, Technische Universit\"{a}t Darmstadt, Darmstadt, Germany
\item \Idef{100}Yonsei University, Seoul, South Korea
\item \Idef{101}Zentrum f\"{u}r Technologietransfer und Telekommunikation (ZTT), Fachhochschule Worms, Worms, Germany
\item \Idef{102}California Polytechnic State University, San Luis Obispo, California, United States
\item \Idef{103}China Institute of Atomic Energy, Beijing, China
\item \Idef{104}Institute of Physics, Academy of Sciences of the Czech Republic, Prague, Czech Republic
\item \Idef{105}University of Tennessee, Knoxville, Tennessee, United States
\item \Idef{106}Dipartimento di Fisica dell'Universit\`{a} `La Sapienza' and Sezione INFN, Rome, Italy
\item \Idef{107}Hiroshima University, Hiroshima, Japan
\item \Idef{108}Lawrence Livermore National Laboratory, Livermore, California, United States
\item \Idef{109}Budker Institute for Nuclear Physics, Novosibirsk, Russia
\item \Idef{110}Physics Department, University of Rajasthan, Jaipur, India
\item \Idef{111}Purdue University, West Lafayette, Indiana, United States
\item \Idef{112}Centre de Calcul de l'IN2P3, Villeurbanne, France 
\item \Idef{113}Pusan National University, Pusan, South Korea
\end{Authlist}
\endgroup